\documentclass[pre,twocolumn,superscriptaddress,floatfix,showpacs]{revtex4}

\usepackage[latin2]{inputenc}
\usepackage{amsmath}
\usepackage{amssymb}
\usepackage{epsfig}
\usepackage{graphicx}

\newcommand{\Sb}{\boldsymbol \sigma}
\newcommand{\Sij}{\sigma_{_{ij}}}

\newcommand{\Saa}{\sigma_{_{11}}}
\newcommand{\Sab}{\sigma_{_{12}}}
\newcommand{\Sba}{\sigma_{_{21}}}
\newcommand{\Sbb}{\sigma_{_{22}}}

\newcommand{\Eb}{\boldsymbol \epsilon}

\newcommand{\Ekl}{\epsilon_{_{kl}}}

\newcommand{\C}{\mathcal{C}}
\newcommand{\Cijkl}{\mathcal{C}_{_{ijkl}}}
\newcommand{\Caaaa}{\mathcal{C}_{_{1111}}}

\newcommand{\Caabb}{\mathcal{C}_{_{1122}}}
\newcommand{\Cabaa}{\mathcal{C}_{_{1211}}}

\newcommand{\Cabbb}{\mathcal{C}_{_{1222}}}
\newcommand{\Cbaaa}{\mathcal{C}_{_{2111}}}

\newcommand{\Cbabb}{\mathcal{C}_{_{2122}}}
\newcommand{\Cbbaa}{\mathcal{C}_{_{2211}}}

\newcommand{\Cbbbb}{\mathcal{C}_{_{2222}}}

\newcommand{\lvec}{\boldsymbol l^c}
\newcommand{\nvec}{\boldsymbol n^c}
\newcommand{\tvec}{\boldsymbol t^c}
\newcommand{\kn}{k_{_n}}
\newcommand{\kt}{k_{_t}}

\begin{document}

\title{Unilateral interactions in granular packings: A model for the anisotropy modulus}

\date{\today}

\author{M.\ Reza Shaebani}
\email{reza.shaebani@uni-due.de}
\affiliation{Department of Theoretical Physics, University of
Duisburg-Essen, 47048 Duisburg, Germany}

\author{Jens Boberski}
\affiliation{Department of Theoretical Physics, University of
Duisburg-Essen, 47048 Duisburg, Germany}

\author{Dietrich E.\ Wolf}
\affiliation{Department of Theoretical Physics, University of
Duisburg-Essen, 47048 Duisburg, Germany}

\begin{abstract}
Unilateral interparticle interactions have an effect
on the elastic response of granular materials due to
the opening and closing of contacts during quasi-static shear
deformations. A simplified model is presented, for which  
constitutive relations can be derived.
For biaxial deformations the elastic behavior in this model 
involves three independent elastic moduli: bulk,
shear, and anisotropy modulus. The bulk and the shear modulus,
when scaled by the contact density, are independent of the 
deformation. However, the magnitude of the anisotropy modulus 
is proportional to the ratio between shear and volumetric strain. 
Sufficiently far from the jamming transition, when corrections due to
non-affine motion become weak, 
the theoretical predictions are qualitatively in agreement with 
simulation results.
\end{abstract}


\maketitle

\section{Introduction}
\label{Introduction}

Understanding the mechanical response of granular materials is still 
one of the remaining challenges in materials science and physics. The 
research is motivated by many industrial and geophysical applications 
\cite{Neddermann92,Jaeger96}. A granular packing at rest does not 
behave like an ordinary elastic solid, because the relation between 
stress and strain is nonlinear, depends on the fabric and on the 
loading path, and gives rise to energy dissipation 
\cite{Elata96,Walton87,Norris97}. These properties have been 
investigated for packings of spheres, where they could be traced back 
to the nonlinearity of Hertzian contacts, disorder, and Coulomb 
friction law. In these studies the contacts between the spheres 
were supposed to remain closed under the strain.

However, there is another source of the nonlinear elastic response in 
granular media, which has not been studied in such detail yet
\cite{Goldenberg2005}. Due to 
the absence of attractive contact forces in dry granular media, the 
contacts can open, and thus do not transmit any elastic restoring nor 
frictional force. Even in the case of linear (Hookean) elasticity on 
the particle level (e.g.\ for packings of parallel cylinders), the 
vanishing of the elastic response under tension renders the macroscopic 
elastic behavior nonlinear. It is this nonlinearity which we analyze 
in this paper. 

It is known that the contact and force networks in sheared granular 
materials are anisotropic \cite{Kruyt04,Rothenburg89,Radjai98}. The 
present analytical approach links the fabric anisotropy to the shear
deformation for fixed volumetric strain. 

In the following, we investigate the incremental linear response 
of a pre-strained two dimensional packing of disks and 
relate the elements of the stiffness tensor to the mean packing 
properties and the probability distribution of contact orientations. 
Then, analytical expressions for the elastic moduli of isotropic and 
sheared contact networks are derived, and the results are compared 
with numerical simulations.

\section{Linear elastic response}
\label{LinearElasticity}
The elastic response of solids to deformations is 
classically expressed in terms of the relationship between the 
{\it stress} tensor $\Sb$ and the {\it strain} tensor $\Eb$. 
Let us assume that the relation between 
the incremental stress tensor $\delta\Sb$ and the incremental 
strain tensor $\delta\Eb$ can be expressed by analytical 
functions $F_{_{ij}}$ as
\begin{equation}
\delta\Sij=F_{_{ij}}(\delta\Ekl) \,,
\label{stress-strain1}
\end{equation}
where $F_{_{ij}}$ satisfy $F_{_{ij}}(\delta \Ekl {=} 0, \hspace{1.3mm} 
\!\! \forall \, k,l \in \{1,. . .,d\}) {=} 0$. Here, $d$ is the 
dimension of the system. In the limit of small deformations, 
Eq.~(\ref{stress-strain1}) can be approximated by Taylor expansion 
to first order in the strain increment
\begin{equation}
\delta\Sij{=}\sum_{k,l} \Cijkl \, \delta\Ekl \,,
\label{stiffness-definition}
\end{equation}
where $\C$ is a tensor of rank $4$, referred to as the {\it stiffness} 
or {\it elasticity} tensor. Equation~(\ref{stiffness-definition}) can 
be regarded as the generalization of {\it Hooke's law} that describes 
the linear response of the medium to external perturbations. Although 
$\C$ has $16$ elements in a two-dimensional system, symmetry considerations for 
the stress and strain tensors imply that there are usually less 
independent elements. For example, the symmetry of a completely 
isotropic material requires that $\C$ has only two independent 
elements, commonly represented by the {\it Lam\'e} coefficients or, 
alternatively, by {\it Poisson's ratio} and {\it Young's modulus}.

In this section, it is recalled, how an approximate expression for the
stiffness tensor of a dense assembly of grains can be obtained, which
allows to calculate the elastic moduli from mean packing
properties and the probability distribution of contact orientations
\cite{Mehrabadi82,Bathurst88}.

\begin{figure}[t]
\centering
\includegraphics[scale=0.3,angle=0]{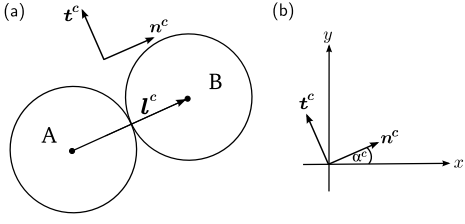}
\caption{(a) Geometry of the contact $c$ between the particles $A$ 
and $B$ including the branch vector $\lvec$, the normal unit vector 
$\nvec$, and the tangential unit vector $\tvec$. (b) $\nvec$ and 
$\tvec$ in an arbitrarily chosen Cartesian coordinates $x{-}y$.}
\label{fig:Schematic}
\end{figure}

The schematic figure \ref{fig:Schematic}(a) shows the undeformed shapes of two particles 
at positions A and B. Their overlap is a measure of the elastic deformation at their 
contact $c$. The normal and tangential contact unit 
vectors are denoted by $\nvec$ and $\tvec$, and the branch vector 
$\lvec$ connects the centers of the particles. Each contact force 
is modeled by two linear springs in the normal and tangential 
directions with the spring constants $k_{_n}$ and $k_{_t}$, 
respectively. This harmonic force law approximates   
the interaction for two-dimensional disks for small 
deformations \cite{Landau86,Johnson87}. Starting from an arbitrary 
weakly deformed state, the change in the force exerted by particle 
B on particle A due to an additional small deformation of the system 
is
\begin{equation}
\delta \boldsymbol f^c =  \kn (\delta\lvec\cdot \nvec) \nvec +\kt 
(\delta\lvec\cdot\tvec) \tvec \,,
\label{contact-force}
\end{equation}
where $\delta \lvec$ is the change of the branch vector due to the 
displacement of the particle centers. It is supposed that the 
contacts do not break due to the imposed deformation. This is 
justified by the fact that during the measurement of the macroscopic 
elastic moduli, the material is subjected only to incremental strain 
changes so that the fabric remains nearly unchanged. The average 
stress increment can be approximated by \cite{Christoffersen81,Luding04}
\begin{equation}
\delta \Sij\approx\frac 1 {V}\sum_{c=1}^{N_c}\delta\!f_i^c\,l_j^c\,,
\label{eq:stress}
\end{equation}
where terms of order (overlap/particle radius) have been neglected. This 
is sufficiently accurate for small pre-strains. Here, the 
sum runs over all contacts $N_c$ within the volume $V$. Using 
Eqs.~(\ref{contact-force}) and (\ref{eq:stress}) one obtains
\begin{equation}
\delta\Sij{=}\frac 1 {V}\sum_{c=1}^{N_c} \left(\kn (\sum_{k} \delta 
l_{_{\!k}}^c \, n_{_{\!k}}^c) n_{_{\!i}}^c {+} \kt (\sum_{k} \delta 
l_{_{\!k}}^c \, t_{_{\!k}}^c) t_{_{\!i}}^c \right) l_{_{\!j}}^c \,.
\label{eq:stress-tensor-2}
\end{equation}
A crucial simplification of this expression is obtained, if one assumes 
{\it affine} displacements of the particle centers, which states 
that the displacement gradient is constant or varies only slowly on 
the scale of the particle size. In this way, the non-affine parts of 
the relative displacements are neglected and the remaining part 
according to the macroscopic strain increment is given by
\begin{equation}
\delta l^c_{_{\!k}} = \sum_{l} \delta\epsilon_{_{kl}} l^c_{_l}\,.
\label{eq:affine}
\end{equation}
Inserting Eq.~(\ref{eq:affine}) into Eq.~(\ref{eq:stress-tensor-2}), 
one obtains
\begin{equation}
\delta\Sij= \sum_{k,l} \frac {1}{V}\sum_{c=1}^{N_c} 
|\lvec|^2\left(\kn n_{_{\!i}}^c n_{_{\!j}}^c n_{_{\!k}}^c n_{_{\!l}}^c +\kt 
t_{_{\!i}}^c n_{_{\!j}}^c t_{_{\!k}}^c n_{_{\!l}}^c \right) \delta
\Ekl \,.
\label{eq:stress-affine}
\end{equation}
For narrow size distributions and small pre-strain, one may replace
$|\lvec|$ by the average particle diameter $\ell$.
Polydispersity leads, in first 
order, to an additional factor that only depends on the moments of 
the particle size distribution \cite{ShaebaniMahyar11}. By comparing 
Eqs.~(\ref{eq:stress-affine}) and (\ref{stiffness-definition}) 
the stiffness tensor is identified as \cite{Liao97,Kruyt98}
\begin{equation}
\Cijkl=\frac {\ell^{^2}}{V}\sum_{c=1}^{N_c} \left(\kn n_{_{\!i}}^c 
n_{_{\!j}}^c n_{_{\!k}}^c n_{_{\!l}}^c +\kt t_{_{\!i}}^c n_{_{\!j}}^c 
t_{_{\!k}}^c n_{_{\!l}}^c \right) \,.
\label{stiffness-tensor-1}
\end{equation}

Denoting the normal and tangential unit vectors as $\vec n{=}
(\cos\alpha , \sin\alpha$) and $\vec t{=}(-\sin \alpha , \cos 
\alpha$) [see Fig.~\ref{fig:Schematic}(b)], the sum over
the contacts located inside  
the volume $V$ can be replaced by an integral over the contact orientation 
distribution $P(\alpha)$ \cite{Bathurst88} so that 
\begin{equation}
\Cijkl{=}\frac{2}{\pi} z \phi \! \int_{-\pi}^{\pi}\!\!\! \left(\kn n_{_{\!i}} 
n_{_{\!j}} n_{_{\!k}} n_{_{\!l}} +\kt t_{_{\!i}} n_{_{\!j}} t_{_{\!k}} 
n_{_{\!l}} \right) P(\alpha)\,d\alpha\,,
\label{eq:elastictensor2}
\end{equation}
where $z\:({=}2N_c/N)$ is the average coordination number, $\phi\;({=} N 
\pi \ell^{^2} / 4V)$ is the volume fraction of the packing, and 
$P(\alpha) \Delta \alpha$ denotes the probability to find a contact 
with an orientation between $\alpha$ and $\alpha+\Delta\alpha$. 
According to Eq.~(\ref{eq:elastictensor2}), the elements of the stiffness 
tensor are determined by the fabric properties: $z$, $\phi$, and $P(\alpha)$.

It is convenient to choose the 
principal axes of the strain tensor increment $\delta \Eb$
as coordinate system so that
\begin{equation}
\delta \Eb = \frac{1}{2}
\left( \begin{array}{ccccccccc}
\delta\epsilon_v{+}\delta\gamma & 0 \\
0 & \delta\epsilon_v{-}\delta\gamma \end{array} \right)
\hspace{6mm} \text{with} \;\; \delta\gamma \geq 0\, ,
\label{strain-principal}
\end{equation}
with $\delta\epsilon_v$(${=}\text{Tr} \delta\epsilon{=}\delta 
V/V$) and $\delta\gamma$, the change in the volumetric strain 
and in the shear deformation, respectively. Then 
Eq.~(\ref{stiffness-definition}) can be written as
\begin{equation}
\left( \begin{array}{ccccccccc}
\delta\Saa \\ \delta\Sbb \\ \delta\Sab \\
\delta\Sba \end{array} \right){=}
\left( \begin{array}{ccccccccc}
\Caaaa & \Caabb \\
\Cbbaa & \Cbbbb \\
\Cabaa & \Cabbb \\
\Cbaaa & \Cbabb \end{array} \right)
\left( \begin{array}{ccccccccc}
(\delta\epsilon_v{+}\delta\gamma)/2 \\ (\delta\epsilon_v{-} 
\delta\gamma)/2 \end{array} \right)\,.
\label{stress-stiffness-strain}
\end{equation}

The balance of torques requires that the average stress 
tensor remains symmetric, i.e. $\delta\Sab{=}\delta\Sba$. Hence, 
the following constraints must hold in equilibrium
\begin{eqnarray}
\Cabaa{=}\Cbaaa \;, \;\;\;\; \Cabbb{=}\Cbabb \,.\,
\label{stress-symmetry}
\end{eqnarray}
Using (\ref{eq:elastictensor2}) this implies that only those
incremental deformations do not induce torques inside the packing (and hence
lead to particle rotations), whose principal axes are compatible with the
contact orientation distribution in the sense that
\begin{equation}
\langle \sin{2\alpha} \rangle = \int_{-\pi}^{\pi}\sin{2\alpha}\ P(\alpha) d\alpha =0.
\label{eq:no_torque_condition}
\end{equation}

Evaluating (\ref{eq:elastictensor2}) for fabrics
that fulfill the constraint (\ref{eq:no_torque_condition}) gives the following
expressions for the elements of the stiffness tensor:
\begin{eqnarray}
\Caaaa &=&\frac{z\phi}{\pi}\left(\kn\left(1+\langle \cos{2\alpha}\rangle\right) 
-\frac{\kn{-}\kt}{2}\langle(\sin{2\alpha})^2\rangle\right)
\label{eq:Caaaa}\\
\Cbbbb &=&\frac{z\phi}{\pi}\left(\kn\left(1-\langle \cos{2\alpha}\rangle\right)
-\frac{\kn{-}\kt}{2}\langle(\sin{2\alpha})^2\rangle\right)
\label{eq:Cbbbb}\\
\Caabb &=& \Cbbaa = 
\frac{z\phi}{\pi}\frac{\kn{-}\kt}{2}\langle(\sin{2\alpha})^2\rangle 
\label{eq:Caabb}\\
\Cabaa &=& \Cbaaa = -\Cabbb = -\Cbabb \nonumber\\
       &=& \frac{z\phi}{\pi}(\kn{-}\kt)\langle \sin{4\alpha}\rangle
\label{eq:Cabaa}
\end{eqnarray}

\section{Biaxial deformations}

In the following we consider biaxial deformations, which means that
the principal axes of the strain tensor do not change, while the
sample is deformed. If the deformation starts from an isotropic
configuration, the fabric ($P(\alpha)$) will remain symmetric with
respect to $\alpha=0$. Hence (\ref{eq:no_torque_condition}) is fulfilled 
and (\ref{eq:Cabaa}) vanishes, as one
averages odd functions with an even distribution. This simplifies
matters considerably, as it implies that the principal axes of stress
and strain coincide so that
\begin{equation}
\delta \Sb = 
\left( \begin{array}{cc}
\delta\sigma_{11} &       0           \\
        0         & \delta\sigma_{22} \\   
\end{array}\right)=-
\left( \begin{array}{ccccccccc}
\delta P+\delta\tau & 0 \\
0 & \delta P-\delta\tau \end{array} \right)\,,
\label{stress-principal}
\end{equation}
where $\delta P$ and $\delta\tau$ are the incremental  
pressure and shear stress. 

One defines the bulk modulus $E$, the shear modulus $G$ and the anisotropy modulus $A$ by
\begin{equation}
\left( \begin{array}{ccccccccc}
\delta P \\
\delta \tau \end{array} \right)=-
\left( \begin{array}{ccccccccc}
E & A \\
A & G \end{array} \right)
\left( \begin{array}{ccccccccc}
\delta\epsilon_v \\
\delta\gamma \end{array} \right).
\label{eq:Definition_of_Moduli}
\end{equation}
According to (\ref{stress-stiffness-strain}) and (\ref{eq:Caaaa}) - (\ref{eq:Caabb})
\begin{eqnarray}
E&=&\frac{\Caaaa+\Caabb+\Cbbaa+\Cbbbb}{4}= \frac{z \phi}{2\pi} \kn, 
\nonumber\\ 
G&=&\frac{\Caaaa-\Caabb-\Cbbaa+\Cbbbb}{4}\nonumber\\
 &&\hspace{1cm} = E 
\left(\langle (\cos{2\alpha})^2\rangle + \frac{\kt}{\kn}\langle (\sin{2\alpha})^2\rangle\right) 
\nonumber\\
A&=&\frac{\Caaaa-\Cbbbb}{4}= E\langle \cos{2\alpha}\rangle . 
\label{eq:Unilateral-Stiffness-6}
\end{eqnarray}

For an isotropic fabric, $P(\alpha)=\frac{1}{2\pi}$, the result of Kruyt and Rothenburg 
\cite{Kruyt98} is reproduced:
\begin{equation}
E=\frac{z \phi}{2\pi} k_n , \quad G=\frac{z \phi}{4\pi} (k_n+k_t) , \quad A=0.
\label{isotropic_case}
\end{equation}
In the next section a simple model is proposed for the kind of anisotropy, a granular 
packing develops under the influence of shear.

\section{Anisotropy induced by unilaterality}
\label{UnilateralElasticity}

In order to elucidate the effect of unilaterality on the stress-strain 
relationship, we must consider the closing respectively opening of
contacts, as a finite volumetric strain $\epsilon_v$ and shear
deformation $\gamma$ build up, starting from an isotropic,
stress-free, jammed packing (unstrained reference state with 
zero overlap). Assuming again that the particle displacements may be 
approximately regarded as affine, the distance between the centers 
of neighboring particles changes due to the strain $\Eb$ by 
\begin{equation}
\Delta\xi_n (\alpha,\Eb) {=} -\ell\sum_{i,j}\epsilon_{ij}n_i n_j 
            {=} -\ell\left(\frac{\epsilon_v}{2}{+}\frac{\gamma}{2}\cos{2\alpha}\right).
\label{eq:Delta_xi}
\end{equation}
The change depends on $\Eb$ as well as on the direction of the branch vector, $\alpha$.
If the two particles touched each other, i.e. $\xi_n=0$ in the
unstrained state, a positive $\Delta\xi_n$ means that the deformation
leads to an overlap, while a negative value indicates that the
particles are no longer in contact in the strained state. 
When there is a gap between the particle surfaces in the unstrained
configuration, an overlap can also form, if $\Delta\xi_n$ is larger 
than the gap.
Therefore we extend the notion of an overlap to include small negative
values, $\xi_n < 0$, which tell the size of the gap.

We introduce the probability density $Q(\xi_n,\alpha,\Eb)$ that 
a particle pair has a branch vector at an angle $\alpha$ with
respect to the principal axis of the strain tensor $\Eb$, 
belonging to the eigenvalue $\epsilon_v+\gamma$, and
that the overlap respectively the negative gap has a value $\xi_n$. 
This probability density depends on the strain $\Eb$.
For $\Eb=0$ it is assumed to be isotropic, i.e. 
independent of $\alpha$, and it fulfills
$Q_0(\xi_n)=0$ for $\xi_n>0$, as there are no overlaps in the unstrained configuration,
and $Q_0(\xi_n)\neq 0$ for $\xi_n\leq 0$, as the unstrained configuration is jammed.
We assume that the probability distribution simply shifts by $\Delta\xi_n$,
Eq. (\ref{eq:Delta_xi}), under the influence of strain:
\begin{equation} 
Q(\xi_n,\alpha,\Eb) = Q_0(\xi_n - \Delta\xi_n(\alpha,\Eb))
\label{eq:Q-model}
\end{equation}
 
The probability that a pair of neighbor particles is actually in contact is
\begin{equation} 
\mathcal{N}=\int_{-\pi}^{\pi} d\alpha \int_0^{\infty} d\xi_n \ Q(\xi_n,\alpha,\Eb)
\label{eq:normalization}
\end{equation}
The probability density, that a contact has a certain angle $\alpha$
with respect to the principal axis of the strain, which belongs to the
eigenvalue $\epsilon_v+\gamma$, is 
\begin{equation} 
P(\alpha,\Eb) =\frac{1}{\mathcal{N}} \int_0^{\infty} d\xi_n \ Q(\xi_n,\alpha,\Eb) .
\label{eq:P_from_Q}
\end{equation}
Applying the assumption (\ref{eq:Q-model}), the integral is in first
order of $\Eb$ given by 
\begin{eqnarray}
\int_0^{\infty} d\xi_n \ Q(\xi_n,\alpha,\Eb) &\approx& 
Q_0(0)\Delta\xi_n(\alpha,\Eb) \nonumber\\
&=&- Q_0(0) \ell \left(\frac{\epsilon_v}{2}+
\frac{\gamma}{2}\cos{2\alpha} \right).
\label{eq:enumerator_first_order}
\end{eqnarray}
Integrating this over $\alpha$ gives the corresponding first order approximation of 
$\mathcal{N}$: 
\begin{equation} 
\mathcal{N} \approx - Q_0(0) \ell\epsilon_v \pi.
\label{eq:N_first_order}
\end{equation}
Hence the properly normalized first order approximation of the
probability density of contact directions is 
\begin{equation} 
P(\alpha,\Eb) \approx \frac{1}{2\pi}\left(1+\frac{\gamma}{\epsilon_v}\cos{2\alpha}\right).
\label{eq:P_first_order}
\end{equation}
This approximation can at most be applied for $\left|\frac{\gamma}{\epsilon_v}\right|\leq 1$, 
as otherwise the probability density would not be positive
semi definite. $\alpha=0$ is the direction of the principal axis
belonging to the eigenvalue $\epsilon_v+\gamma$. For $\epsilon_v<0$
(compressive strain) and $\gamma \ge 0$ (by definition) this is the
direction in which the precompressed system expands during biaxial deformation.
Therefore contacts preferentially open in this direction so that
$P(0,\Eb) \approx \frac{1}{2\pi}\left(1+\frac{\gamma}{\epsilon_v}\right) 
< \frac{1}{2\pi}$.

Equation (\ref{eq:P_first_order}) is the main new result of this
paper. In this approximation,
$\langle(\cos{2\alpha})^2\rangle=\langle(\sin{2\alpha})^2\rangle=\frac{1}{2}$
and $\langle\cos{2\alpha}\rangle=\frac{\gamma}{2\epsilon_v}$ so that
the elastic moduli of a granular packing are approximately
\begin{equation} 
E=\frac{z \phi}{2\pi} k_n , \quad G=\frac{z \phi}{4\pi} (k_n+k_t) , 
\quad A=\frac{z\phi}{4\pi}\kn\frac{\gamma}{\epsilon_v}.
\label{eq:anisotropic_case}
\end{equation}
Note that due to the presence of the nonzero element $A$ in 
Eq.~(\ref{eq:Definition_of_Moduli}), two independent 
experimental tests are required to determine the elastic moduli 
of an anisotropic material, for example:
\begin{enumerate}
\item[(I)] \emph{incremental $\delta \epsilon_v$ while $\delta \gamma {=}0$}:
\begin{equation}
E=-\left.\frac{\delta P}{\delta \epsilon_v}\right|_{\delta \gamma {=}0},
\;\;\;\;\;\;\;
A=-\left.\frac{\delta \tau}{\delta \epsilon_v}\right|_{\delta \gamma {=}0}.
\label{experiment1}
\end{equation}
\item[(II)] \emph{incremental $\delta \gamma$ while $\delta \epsilon_v {=}0$}:
\begin{equation}
G=-\left.\frac{\delta \tau}{\delta \gamma}\right|_{\delta \epsilon_v {=}0},
\;\;\;\;\;\;\;
A=-\left.\frac{\delta P}{\delta \gamma}\right|_{\delta \epsilon_v {=}0}.
\label{experiment2}
\end{equation}
\end{enumerate}
This is in contrast to the isotropic case, where a single experiment 
with simultaneous incremental $\delta \epsilon_v$ and $\delta \gamma$ is 
sufficient to measure both bulk and shear moduli. In granular media, in 
contrast to an isotropic elastic material, a pure shear leads to a 
pressure increase, $\delta P = -A\delta\gamma > 0$ as $A<0$.

\section{Simulation results}
\label{SimulationResults}

We tested the theoretical predictions of Sec.\ref{UnilateralElasticity}
by numerical simulations.
The unstrained initial packing consists of 3000 rigid disks 
with particle radii uniformly distributed between $a_\text{min}{=}0.95$ 
and $a_\text{max}{=}1.05$ to avoid crystalline order. It was 
generated by a method based on Contact Dynamics simulations, 
which leads to homogeneous, isotropic, jammed configurations
\cite{Shaebani09}. Such a packing is taken as unstrained initial
configuration in a Molecular Dynamics simulation of soft particles using
the LAMMPS code \cite{Plimpton95,LAMMPS}.
The interactions between particles are modeled with 
normal and tangential Hookean springs (with $k_t/k_n{=}0.5$).
We checked that the results do not depend on the friction coefficient,
when it is chosen larger than 0.5. The data shown here are for $\mu=1$.  

\begin{figure}[t]
\centering
\includegraphics[scale=1]{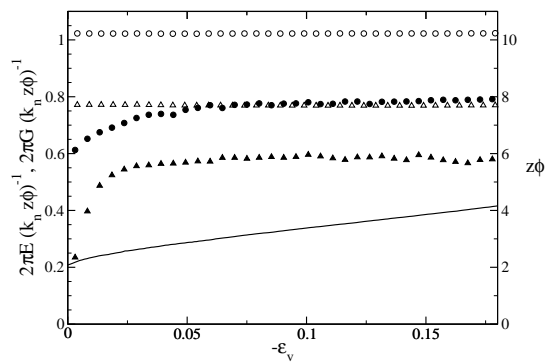}
\caption{Bulk respectively shear modulus (circles respectively
  triangles), determined during an isotropic compression, in which the
  negative volumetric strain $-\epsilon_v$ is increased in small
  steps and $\gamma=0$. Both moduli are given in units 
  of $\kn/(2\pi)$ and divided by $z\phi$. They were
  calculated from fabric (method 1, open symbols), respectively 
  from Eq.(\ref{experiment1})(method 2, full symbols).
  The full line represents $z\phi$, referring to the scale
  on the right.}  
\label{fig:E_and_G_vs_epsilon}
\end{figure}
 
In a quasi-static compression 
process, the volume of the unstrained packing is gradually 
decreased by applying incremental volumetric strain steps and 
allowing the system to relax between those steps. 
The process is stopped 
when the total volumetric strain $\epsilon_v=\Delta V/V$ reaches 
a given value ($\epsilon_v = -0.04$ respectively $\epsilon_v = -0.09$).
This precompressed packing is then sheared in a biaxial geometry,
while keeping the volume of the system constant. 
The shear deformation is imposed via 
incremental steps, and the system is allowed to 
relax between the steps. 

\begin{figure}[b]
\centering
\includegraphics[scale=1]{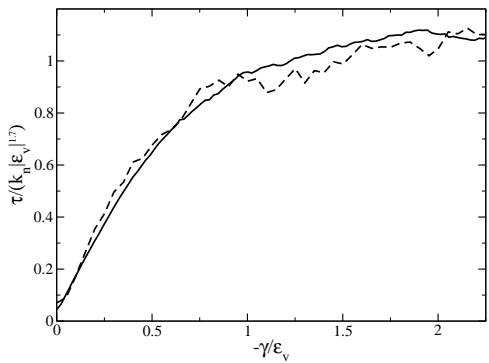}
\caption{Shear stress in unites of $\kn$, scaled by $|\epsilon_v|^{1.7}$
  as a function of shear strain 
  $\gamma$ scaled by $-\epsilon_v$. Dashed line: $\epsilon_v=-0.04$,
  full line: $\epsilon_v=-0.09$.}
\label{fig:elastic_regime}
\end{figure}
 
Two different methods are used to determine the elastic moduli:
\begin{enumerate}
\item Using Eq. (\ref{stiffness-tensor-1}) the elastic constants can
  be computed from the fabric. This formula assumes affine
  deformations and that the contact network remains unchanged for
  incremental strains.
\item An incremental strain test is simulated by molecular dynamics
  and the elastic moduli are determined from Eqs. (\ref{experiment1})
  and (\ref{experiment2}). This method allows for a change of the
  contact network and non-affine motions of the particles.
\end{enumerate}

\begin{figure}[t]
\centering
\includegraphics[scale=1]{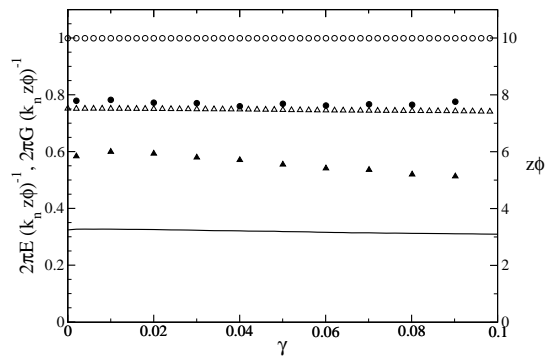}
\caption{Same as Fig.\ref{fig:E_and_G_vs_epsilon}, but now as function
  of shear strain $\gamma$ for fixed $\epsilon_v=-0.09$.}
\label{fig:E_and_G_vs_gamma}
\end{figure}

Fig.\ref{fig:E_and_G_vs_epsilon} shows that the bulk and shear moduli
divided by $z\phi$ are approximately constant for
strong enough compression, $-\epsilon_v>0.04$, in agreement with
Eq.(\ref{eq:anisotropic_case}). However, as one
approaches the unstrained configuration (at the jamming transition),
the elastic moduli obtained from evaluating incremental strain tests
soften, whereas the ones calculated from the fabric remain
unchanged. This shows that close to the jamming transition the
assumption that the incremental particle movements are affine fails, in
agreement with the findings of \cite{Heussinger2009,vanHecke2010}. 
Therefore we concentrate on
the regime $-\epsilon_v\geq 0.04$ in the following. Remarkably, 
the ratio of the two moduli, $G/E$ is very close to the prediction of
the simple theory, $G/E = (1 + \kt/\kn)/2 = 0.75$. Also, the values 
calculated from the fabric agree with the theory, whereas the ones
obtained by evaluating Eqs.(\ref{experiment1}) and (\ref{experiment2})
are about 20 \% smaller.

Fig.\ref{fig:E_and_G_vs_epsilon} also shows that the contact density
($= \frac{2}{\pi \ell^2}z\phi$)
increases the more compressed the packing is, while it decreases, if a
precompressed sample is sheared at fixed volume, see
Fig.\ref{fig:E_and_G_vs_gamma}, provided $\gamma$ remains small
enough. For large shear deformation, the system presumably forms a
shear band and begins to flow. Then the contact density must become
independent of $\gamma$. The transition from elastic to plastic
response can be seen in Fig.\ref{fig:elastic_regime}. The shear stress
first increases linearly with $\gamma$ and saturates for large
deformation, indicating the plastic regime. This shows, that one can
speak about elastic response only for $\frac{-\gamma}{\epsilon}<1$.
  
\begin{figure}
\centering
\includegraphics[scale=1]{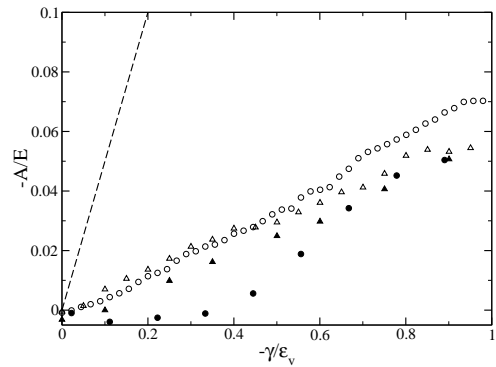}
\caption{Ratio of negative anisotropy modulus and bulk modulus,$-A/E$,
  as function of the ratio between shear strain and negative
  volumetric strain, $-\gamma/\epsilon_v$. Data are shown for two
  volumetric strains, $\epsilon_v=-0.04$ (circles) and
  $\epsilon_v=-0.09$ (triangles). They were
  calculated from fabric (method 1, open symbols), respectively 
  from Eq.(\ref{experiment1})(method 2, full symbols). The dashed line
  is the theoretical result, Eq.(\ref{eq:anisotropic_case}).}
\label{fig:A_vs_gamma/epsilon}
\end{figure}

Fig.\ref{fig:E_and_G_vs_gamma} confirms that bulk and shear moduli, divided
by the contact density, do not change, when a precompressed state is
sheared, in agreement with the theoretical result
Eq.(\ref{eq:anisotropic_case}). The model predicts, however, that the
anisotropy modulus is proportional to $\gamma/\epsilon_v$ (dashed line
in Fig.\ref{fig:A_vs_gamma/epsilon}). Indeed, this is
confirmed in the simulation. The results for $A/E$ determined from the
fabric, as well as the ones from the incremental shear tests for 
$\epsilon_v=-0.04$ agree with each other within the error bars and
show can be fitted by a linear dependence on $\gamma/\epsilon_v$ .
Only the shear test simulation
data for $\epsilon_v=-0.09$ deviate from this behavior for small
shear, $-\gamma/\epsilon_v < 0.7$, although the evaluation of the
fabric perfectly agrees. Further simulations are needed to clarify the 
origin of this peculiar behaviour. 
The simulation data essentially confirm the linear dependence of the
anisotropy modulus on the ratio $\gamma/\epsilon_v$. However, the
theory overestimates the slope by a factor of about 7.

\section{Conclusion}
\label{DiscussionsCoclusion}

We have shown that the opening and closing of contacts can explain the
anisotropy modulus which is characteristic for the elastic response
of dense granular packings under biaxial shear. The theory predicts a
linear dependence of 
the anisotropy modulus on the ratio $\gamma/\epsilon_v$, which is the
only zeroth order combination of the scalar invariants
$\gamma = \sqrt{{\rm Tr}\Eb^2-2 \det\Eb}$ and $\epsilon_v={\rm Tr}\Eb$. This is
confirmed by simulations, but the theory overestimates the anisotropy 
modulus by a factor of 7. The bulk and shear moduli are predicted
correctly by the theory, as long as one is not too close to the
jamming transition, where the assumption of affine deformation of the
packing fails.

\section*{Acknowledgment}
\label{acknowledgments}
We would like to thank Isaac Goldhirsch and Tam\'as Unger for fruitful discussions. This work 
was supported by the German Research Foundation (DFG) via priority
program SPP 1486 ``Particles in Contact''.

\end{document}